\documentclass[aps,prc,showpacs,twocolumn,preprintnumbers,floatfix,
showkeys,tightenlines]{revtex4}
\usepackage{amsmath,amssymb,amsfonts}
\usepackage{graphicx}
\usepackage{color}
\allowdisplaybreaks
\begin{document} 

\title{The effects of medium on nuclear properties in multifragmentation}

\author{J. N. De$^{1}$, S. K. Samaddar$^{1}$, X. Vi\~nas$^{2}$, 
M. Centelles$^{2}$, I. N. Mishustin$^{3,4}$, and W. Greiner$^{3}$}
\affiliation{
$^1$Saha Institute of Nuclear Physics, 1/AF Bidhannagar, Kolkata
{\sl 700064}, India \\
$^2$Departament d'Estructura i Constituents de la Mat\`eria,
Facultat de F\'{\i}sica, \\
and Institut de Ci\`encies del Cosmos, Universitat de Barcelona, \\
Diagonal {\sl 645}, {\sl 08028} Barcelona, Spain\\
$^3$Frankfurt Institute for Advanced Studies, J. W. Goethe University,\\
{D-60438}, Frankfurt Am Main, Germany \\
$^4$Kurchatov Institute, Moscow {\sl 123182}, Russia}


\begin{abstract}

In multifragmentation of hot nuclear matter, properties of fragments
embedded in a soup of nucleonic gas and other fragments should be
modified as compared with isolated nuclei. Such modifications are
studied within a simple model where only nucleons and one kind of
heavy nuclei are considered. The interaction between different 
species is described with a momentum-dependent two-body potential
whose parameters are fitted to reproduce properties of cold isolated
nuclei. The internal energy of heavy fragments is parametrized according
to a liquid-drop model with density and temperature dependent parameters.
Calculations are carried out for several subnuclear densities and 
moderate temperatures, for isospin-symmetric and asymmetric systems.
We find that the fragments get stretched due to interactions with
the medium and their binding energies decrease with increasing
temperature and density of nuclear matter. 
\end{abstract}



\maketitle

\section{Introduction}

In energetic nuclear collisions, the participating hot nuclear matter
after an initial dynamic stage of compression expands to a subsaturation
density and then disassembles into many fragments due to growing
instability. Statistical models of different genres \cite{bon1,bon2,gro,
pal,de1,pan} have generally been successful in explaining the many features
associated with the fragment multiplicities, the caloric curve, the 
density of the fragmenting systems, etc. They also offer a broad hint
about the general nature of the phase diagram of nuclear matter \cite
{mul,sil} at temperature $T \sim $ 3$-$8 MeV at subsaturation densities
$\rho \sim $ 1/20 to 1/5th of the normal nuclear density $\rho_0$.
Possible liquid-gas phase transition and associated condensation
\cite{mey,hor,de2} to form nuclear clusters at these temperatures and
densities help in a better exploration of many 
 phenomena of astrophysical interest
like supernova explosions or explosive nucleosynthesis 
\cite{shen,ish,bot1,bot3,hem,fur}.

Analysis of recent laboratory experiments \cite{fev,igl} on nuclear
multifragmentation seems to indicate that the properties of the nuclides
are modified at  subnuclear densities ($\rho \sim \rho_0/3 $)
they are created in corresponding to freeze-out. 
The symmetry energy, for example, is reported
to be progressively reduced \cite{she1} with excitation energy, which
is attributed to the in-medium modifications of the properties of the hot
fragments \cite{sou,ogu}. A looming uncertainty about whether the measured
symmetry energy corresponds to the hot fragments or the fragmenting
system \cite{ono,she1}, however, does not allow an equivocal decision
about the medium modifications. The reduction in symmetry energy 
can have a fair explanation from the thermal and expansion effects 
of the disassembling system \cite{sam1}. The surface properties
of the hot fragments \cite{bot2} as well as their bulk energy \cite{buy}
are also speculated to be modified due to the embedding environment.
 A quantum statistical approach to the nuclear equation of state
taking into account the
formation of clusters  \cite{typ,vos} shows 
that the properties of these clusters are modified due to the
medium in which they are  formed. The symmetry energy of low density
warm nuclear matter predicted by this model seems to be in
good agreement with the experimental data \cite{nat1}.  
In a recent experiment \cite{hag}, it is claimed that the binding energies
of very light clusters ($A \le 4$) produced in multifragmentation
progressively tend to zero in the temperature range of $T \sim $
5 $-$ 10 MeV even at a very low in-medium density $ \sim $ 0.05 $\rho_0$.

The changes, if any, of the bulk properties of the fragments produced in
nuclear disassembly are expected to originate from the effects of the residual
interaction of the fragments with the surroundings. The aim of the
present article is to study these effects starting from an effective
nucleon-nucleon interaction. 
To keep the physics simple and transparent and
yet retaining all the basic essentials, we allow the dilute matter
to condense into only one kind of nuclear species surrounded by a hot
nucleonic gas and species of the same kind. Then we introduce 
the  interactions between them and
look for the minimum of the free energy of the system with variation of the
size of the fragments maintaining chemical equilibrium 
between the fragments and the nucleon gas. The energy and free
energy of the nuclei are evaluated in the liquid-drop framework
that makes it easier to account for the associated changes in the surface and
symmetry energy with the change in the volume of the fragment species.

The organization of the paper is as follows. In Sec. II, the outlines of
the theory are given. Results and discussions are contained in Sec. III.
The concluding remarks are presented in Sec. IV.

\section{Theoretical formulation}

As is well known, hot  low-density nuclear matter 
condenses into nuclear fragments 
of different sizes surrounded by nucleons. We postulate that the 
nucleons and the fragments interact through a common effective 
interaction. The bulk properties of the fragments
may be  modified because of this interaction too. For a
qualitative understanding of such system,  we take only one 
kind of fragment species of mass $A$ and charge $Z$. The interaction is chosen
to be the modified Seyler-Blanchard (SBM) interaction.
Its properties are summarized
in Sec.~IIA. In Sec.~IIB,  the method for 
evaluating  the 
nucleon-fragment  and the fragment-fragment interactions
is described. In Sec.~IIC, we study the observables sensitive to
the medium modification of the properties of finite nuclei.

\subsection{The effective interaction }

 The SBM interaction is a momentum and density dependent effective
interaction of finite range. In the context of the nuclear mass formula,
an interaction of this type has been used with great success by
Myers and Swiatecki \cite{mye}. It also reproduces the rms radii,
charge distributions, and giant monopole resonance energies for
a host of even-even nuclei ranging from $^{16}$O to very heavy
systems \cite{de3}. Its form is given by

\begin{eqnarray}
v(r,p,\rho )&=&C_{l,u}\left [v_1(r,p)+v_2(r,\rho )\right ], \nonumber \\
v_1&=&-(1-\frac{p^2}{b^2})f({\bf r_1,  r_2}), \nonumber \\
v_2&=&d^2\left [ \rho (r_1) + \rho (r_2) \right ]^\kappa f({\bf r_1, r_2}), 
\end{eqnarray}
with 
\begin{eqnarray}
f({\bf r_1, r_2})&=& \frac{e^{-|{\bf r_1-r_2 }|/a}}{|{\bf r_1-r_2 }|/a}.
\end{eqnarray}
 The subscripts $l$ and $u$ to the interaction strength $C$ refer to
like-pair (nn or pp) and unlike-pair (np)
interactions, respectively. The relative separation of the interacting
nucleons is ${\bf r=r_1-r_2 }$ and the relative momentum is
${\bf p=p_1-p_2 }$.
The potential parameters $C_l, C_u, a, b,
d$, and  $\kappa $ are listed in Table I. The procedure for determining
these parameters are given in detail in Ref. \cite{bandy}. These parameters
are somewhat different
from those given in \cite{de3}; in the latter, the symmetry coefficient
$a_{sym}$ for infinite nuclear matter at saturation density $\rho_0$
was chosen to be 34 MeV; in the present
calculation, it is taken to be 31 MeV to be more consistent with the
recent estimates \cite{she2,dan}. The effective mass of the nucleon
coming from the momentum dependence of this effective interaction
is 0.62$m$ for symmetric nuclear matter, where $m$ is the nucleon 
mass. For the interaction, the isoscalar volume incompressibility
$K_{\infty}$, symmetry incompressibility $K_{sym}$ and 
$L$, a measure of the symmetry pressure 
are $240, -101,$ and $59.8$ MeV, respectively.
 It is interesting to note that the symmetry coefficients
$a_{sym}$, $L$, and $K_{sym}$ of this interaction are within the range
of values suggested by the empirical constraints emerging from recent
analysis of different observables 
\cite{she2,dan,chen10,war,roca11,chen12,dong12}.
With this interaction, for symmetric nuclear matter, the critical
temperature is reached at $T_c$ =14.9 MeV, when the surface energy vanishes.

\begin{table}
\caption{ The parameters of the effective interaction (in MeV fm units)}
\begin{ruledtabular}
\begin{tabular}{cccccc}
$C_l$&  $C_u$& $a$& $b$& $d$& $\kappa $\\
\hline
348.5& 829.7& 0.6251& 927.5& 0.879& 1/6\\
\end{tabular}
\end{ruledtabular}
\end{table}

\subsection{Nucleon-fragment and fragment-fragment interaction energy}

A low-density nucleonic matter of density $\rho_b$ and asymmetry
$X=(\rho_n^b-\rho_p^b)/\rho_b$ 
breaks up to a system of free (unbound) neutrons
and protons of density $\rho_n$ and $\rho_p$ and a collection of mass-$A$
fragments (we call it AN matter), all at temperature $T$. 
The clusterized matter is thermodynamically more favorable than the
uniform matter at the same (low) density, asymmetry and temperature \cite{de4}. 
The baryonic
density is then given by
\begin{eqnarray}
\rho_b=\rho_N+A\rho_A,
\end{eqnarray}
where $\rho_N=\rho_n+\rho_p$ is the free nucleonic density and $\rho_A$
is the number density of the fragment species. 
The total thermodynamic potential
$\Omega $ of the system is 
\begin{eqnarray}
\Omega=E-TS-\sum_{\tau} \mu_{\tau} N_{\tau} -\mu_A N_A,
\end{eqnarray}
where $E,S,\mu_{\tau} ,\mu_A, N_{\tau} $,and $N_A$ are the total energy,
entropy, chemical potentials of the free nucleons and the fragments,
free nucleon number and the number of the fragments of mass $A$,
respectively. The isospin index (n,p) is represented by $\tau $. 
Chemical equilibration ensures
\begin{eqnarray}
\mu_A=N\mu_n+Z\mu_p,
\end{eqnarray}
where $N$ and $Z$ are the neutron and proton numbers in the fragment.
The total internal energy of the AN system is written as
\begin{eqnarray}
E=E_{NN}+E_{AN}+E_{AA}.
\end{eqnarray}
In Eq.~(6), $E_{AN}$ is the contribution coming from the nucleon-fragment
interaction $(V_{AN})$. $E_{NN}$ measures the kinetic energy of the
free nucleons plus the interaction energy amongst themselves.  
$E_{AA}$ is the sum total of the the kinetic energy of the fragments, 
interaction energy  among them and their binding energies.

Assuming for simplicity that the fragments are sharp-surface 
liquid drops with a uniform nucleon density $\rho_l$,
these terms can be explicitly written as,
\begin{eqnarray}
E_{NN}&=& \sum_\tau\Biggl  \{\int d{\bf r_1} d{\bf p_1} \frac{p_1^2}{2m_\tau } 
\tilde n_\tau ({\bf p_1}) +\frac{1}{2} \int d{\bf r_1}d{\bf p_1}
d{\bf r_2}d{\bf p_2} \nonumber  \\
&&\times [v_1(|{\bf r_1}-{\bf r_2}|, |{\bf p_1}
-{\bf p_2}|)+v_2(|{\bf r_1}-{\bf r_2}|,2\rho_N)] \nonumber \\
&& \times [C_l\tilde n_\tau ({\bf p_2}) 
+C_u \tilde n_{-\tau }({\bf p_2})] \tilde n_\tau ({\bf p_1})\Biggr \}, 
\end{eqnarray}
\begin{eqnarray}
E_{AN}&=&\frac{1}{2} \sum_{\tau} \Biggl \{
\int d{\bf r_1} d{\bf p_1} d{\bf r_2}
d{\bf p_2} \nonumber \\
&&\times \tilde n_\tau ({\bf r_1,\bf p_1}) \tilde n_A ({\bf r_2,\bf p_2})
\sum_{\tau^\prime} (C_l\delta_{\tau \tau^\prime }
+C_u(1-\delta_{\tau \tau^\prime })) \nonumber \\
&&\times \int_{V_A} d{\bf r}\int d{\bf p}_{l\tau^\prime}^A \tilde 
n_{l\tau^\prime}^A 
({\bf r,\bf p}_{l\tau^\prime}^A) 
\Bigl [v_1(|{\bf {r+R}}|,|{\bf p_1}
\nonumber \\
&&-({\bf p}_{l\tau^\prime}^A+{\bf p_2})|)
+ v_2(|{\bf {r+R}}|,\rho_N 
+\rho_l )
\Bigr ] \Biggr \} 
\end{eqnarray}
and,
\begin{eqnarray}
E_{AA}=E_{AA}^0-N_AB_A(\rho_l ,T),
\end{eqnarray}
where,
\begin{eqnarray}
E_{AA}^0&=& \int d{\bf r}d{\bf p}\frac{p^2}{2m_A} \tilde n_A({\bf r},{\bf p})
+\frac{1}{2} \int d{\bf r_1}d{\bf p_1} d{\bf r_2}d{\bf p_2} \nonumber \\
&&\times \tilde n_A({\bf r_1},{\bf p_1}) \tilde n_A({\bf r_2},{\bf p_2}) 
\int_{V_A}d{\bf r}d{\bf r^\prime } \nonumber \\
&&\times \sum_{\tau ,\tau^\prime}
(C_l\delta_{\tau \tau^\prime}+C_u(1-\delta_{\tau \tau^\prime})) 
\int d{\bf p}_{l\tau}^Ad{\bf p}_{l\tau\prime}^A \nonumber \\
&&\times \tilde n_{l\tau}^A({\bf r},{\bf p}_{l\tau}^A)\tilde n_{l\tau^\prime }^A
({\bf r}^\prime ,{\bf p}_{l\tau^\prime}^A) 
\Bigl [v_1(|{\bf {R+r-r^\prime }}| ,|({\bf p_1} \nonumber \\
&&+{\bf p}_{l\tau}^A) 
 -({\bf p_2}+{\bf p}_{l\tau^\prime}^{A^\prime})|) \nonumber \\
&&+v_2(|{\bf {R+r-r^\prime}}| ,2\rho_l)\Bigr ].
\end{eqnarray}
In Eq.~(8),   
${\bf R=r_2-r_1 }$ is the distance between the nucleon and
the center of the nucleus (see Fig.~1). In Eq.~(10), ${\bf R}$ is the
distance between the two fragment centers (see Fig.~2).
 The various space coordinates occurring
in Eqs.~(8) and (10)
are shown in Figs.~1 and 2, respectively.
In evaluating the coordinate-space integrals in Eqs.~(7), (8) and (10), 
we have assumed,
as in the calculation of the equation of state of dilute nuclear 
matter \cite{shen1,sam5}
that the free nucleons do not penetrate the sharp surface nuclei 
and also that the fragments do not interpenetrate 
so that the identity of the free nucleons and the fragments is never
altered. This 'no-overlap' approximation is tantamount to use of the
'excluded-volume' correction employed earlier \cite{bon2} where the
'free' volume available to fragments is reduced compared to the
total volume $V$ by at least the internal volume of the nucleons
and the fragments. 
\begin{figure}
\includegraphics[width=1.0\columnwidth,angle=0,clip=true]{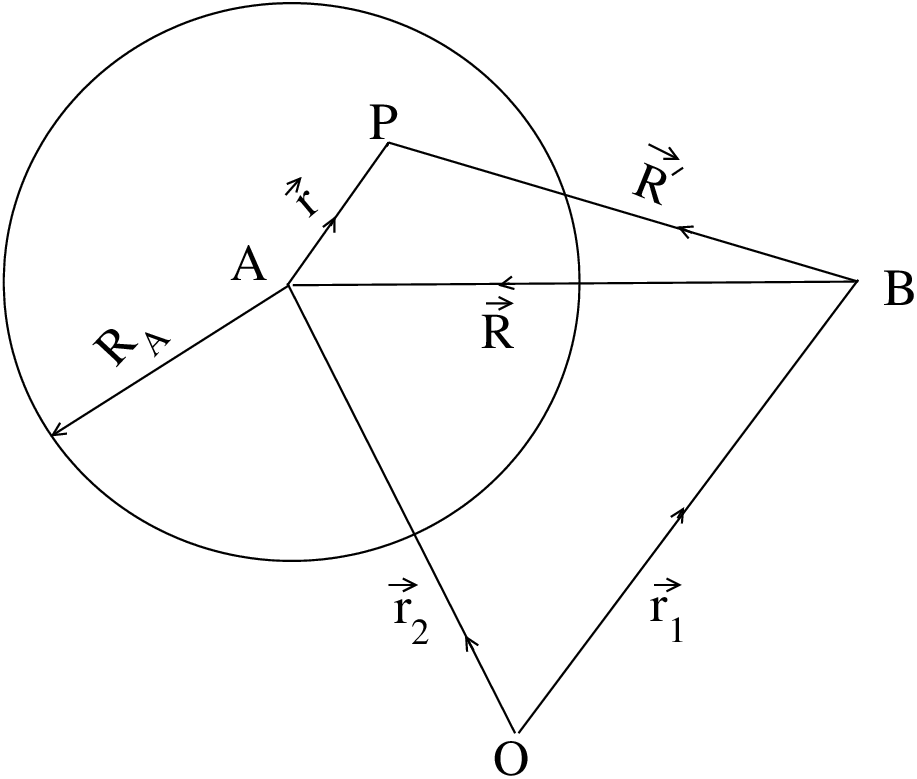}
\caption{ Space coordinates shown
for nucleon (located
at $B$) and fragment (with center at $A$)
configuration. The origin 
of the coordinate system is at $O$
and $P$ is any arbitrary point within the fragment.}
\end{figure}

\begin{figure}
\includegraphics[width=1.0\columnwidth,angle=0,clip=true]{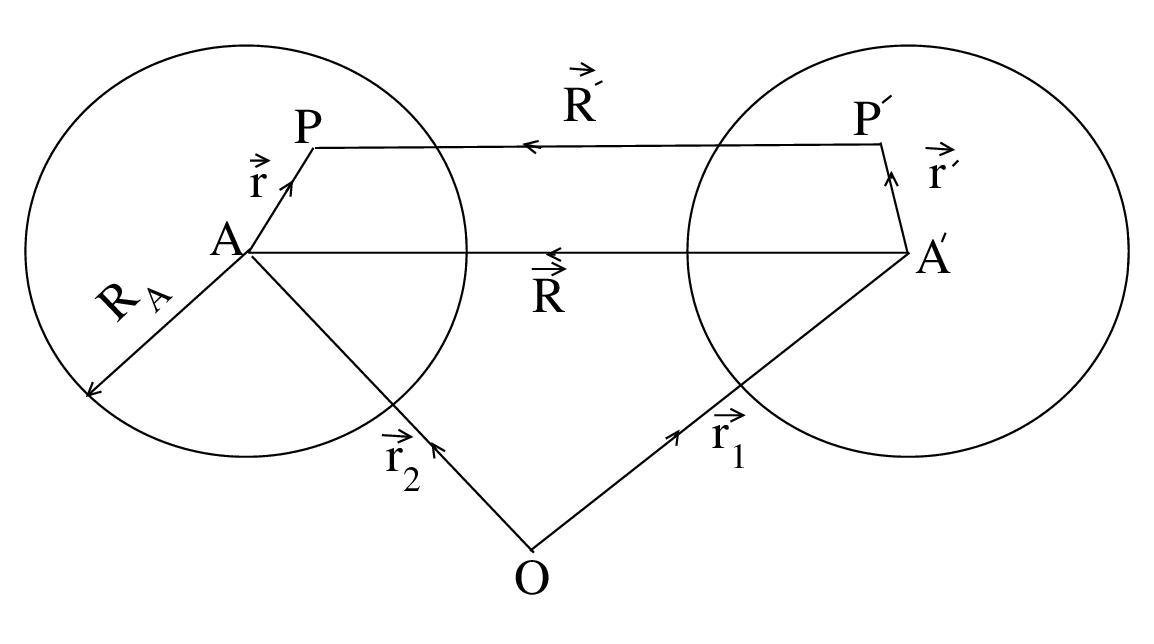}
\caption{ Space coordinates shown for fragment-fragment configuration with
 $O$ as the origin
of the coordinate system. $P$ and
$P^{\prime}$ are arbitrary points within the fragments
with $A$ and $A^{\prime}$ as their centers.}
\end{figure}

In the above equations, $m_\tau $ and $m_A$ are the masses of the nucleons
and the fragments, and $B_A(\rho_l ,T)$ is the binding energy of the produced
fragments at temperature $T$ with internal nucleon  density $\rho_l$. 
Here
$\tilde n_\tau =\frac {2}{h^3} n_\tau $ and $\tilde n_A=\frac 
{g_A}{h^3}n_A$; $n_\tau $ and $n_A$ are the occupation
probabilities for the free nucleons and 
the fragments and $g_A$ is the degeneracy
of the fragments taken to be 1 or 2 depending on whether they are 
bosons or fermions. 
Since the system is infinite, the occupation functions $\tilde n_\tau $  
are independent of space coordinates. 
Similar is the case for $\tilde n_A$.  
In the equations above, space dependence has, however,  
been retained to correlate
with Figs.~1 and 2. Since the fragments are also taken to be uniform 
drops, $\tilde n_{l\tau }^A$ ($=\frac{2}{h^3}n_{l\tau}^A$), 
the distribution function of the
constituent nucleons inside  the fragments is also independent of the
space coordinates. The explicit functional dependence of the
distribution functions on the space coordinates is henceforth
omitted from the equations where the distribution functions may enter.
The momenta of these nucleons inside the
fragment is designated by ${\bf p}_{l\tau}^A$. The notation
$\int_{V_A} $ refers to  the configuration integral over the
volume $V_A$ of the fragment. All other integrals are over the entire 
configuration or momentum space unless otherwise specified.
The distribution functions yield the densities as,
\begin{eqnarray}
\frac{2}{h^3}\int n_\tau ({\bf p})d{\bf p }= N_\tau/V =\rho_\tau,
\end{eqnarray}
\begin{eqnarray}
\frac{2}{h^3}\int n_A ({\bf p})d{\bf p }= N_A/V =\rho_A,
\end{eqnarray}
\begin{eqnarray}
\frac{2}{h^3}\int n_{l\tau}^A ({\bf p})d{\bf p }= A_\tau/V_A =\rho_{l\tau },
\end{eqnarray}
where $V$ is the volume of the AN matter, $A_\tau $ is the neutron
number $N$ or proton number $Z$ in the fragment, 
$\rho_N=\sum_\tau \rho_\tau $, $\rho_l=\sum_\tau \rho_{l\tau }$
and $V_A=\frac {4\pi}{3} R_A^3 $.
In Eq.~(13), $\rho_{l\tau}$ refers to the neutron or
proton number density in the fragment; $R_A$ is its sharp-surface radius.

The total entropy of the AN system is
\begin{eqnarray}
S=\sum_\tau S_\tau +S_A^{tr} +S_A^{int},
\end{eqnarray}
where in the Landau quasiparticle approximation, the contribution
$\sum_\tau S_\tau $ of the free nucleons is taken as 
\begin{eqnarray}
\sum_\tau S_\tau &=&-\frac{2}{h^3}\sum_\tau \int d{\bf r}d{\bf p}
\Bigl [n_\tau ({\bf p})\ln n_\tau ({\bf p}) \nonumber \\
&&+(1-n_\tau ({\bf p}))\ln (1-n_\tau ({\bf p}))\Bigr ].
\end{eqnarray}

$S_A^{tr}$ is the entropy from the center-of-mass motion of the fragments,
and $S_A^{int}$ is their internal entropy. $S_A^{tr}$ is evaluated as,
\begin{eqnarray}
S_A^{tr} &=&-\frac{g_A}{h^3} \int d{\bf r}d{\bf p}
[n_A ({\bf p})\ln n_A ({\bf p}) \nonumber \\
&&\pm (1 \mp n_A ({\bf p}))\ln (1 \mp n_A ({\bf p}))].
\end{eqnarray}
In the above equation, the upper and lower signs correspond to a 
fermionic and bosonic fragment, respectively. 
The contribution $TS_A^{int}$ from internal entropy to the
thermodynamic potential can be absorbed alongwith 
the binding energy term  
of Eq.~(9) in the free energy of the fragments 
$F_A(\rho_l ,T)=(-B_A(\rho_l ,T)-TS_A^{int})$ 
 when the thermodynamic potential
takes the form,
\begin{eqnarray}
\Omega &=&E_{NN}+E_{AN}+E_{AA}^0
 -T \Bigl (\sum_\tau S_\tau +S_A^{tr} \Bigr ) \nonumber \\
&& -\sum_\tau \mu_\tau N_\tau -\mu_AN_A+N_AF_A(\rho_l ,T)
\end{eqnarray}
Minimization of $\Omega $ with respect to $n_\tau $ and $n_A$,
remembering that $\delta n_\tau ({\bf p})$ and $\delta n_A({\bf p})$
are separately arbitrary over the whole phase space, after some algebraic
manipulations, yields for the distribution functions with the 
following structures,
\begin{eqnarray} 
n_\tau ({\bf p})=\left [\exp \left ( \frac{p^2}{2m_\tau^*T}
-\eta_\tau \right ) +1 \right ]^{-1},
\end{eqnarray} 
\begin{eqnarray} 
n_A ({\bf p})=\left  [\exp \left ( \frac{p^2}{2m_A^*T}
-\eta_A  )\right ) 
\pm 1 \right ]^{-1}.
\end{eqnarray} 

In Eqs.~(18) and (19), $\eta_\tau=(\mu_\tau-V_\tau^0-V_\tau^2)/T$
and $\eta_A=(\mu_A-F_A-V_A^0)/T$ are the fugacities pertaining
to the free nucleons and fragments, respectively
and $m_\tau^*$ and $m_A^*$ are the
effective masses of the nucleons and the fragments in the medium,
the masses getting renormalized owing to the momentum dependence of
the force. The nucleonic rearrangement potential $V_\tau^2$ originates
from the density dependence of the interaction.
 The effective nucleon and fragment 
masses are given by
\begin{eqnarray} 
m_\tau^*=\left [\frac{1}{m_\tau}+2V_\tau^1 \right ]^{-1},
\end{eqnarray} 
and
\begin{eqnarray} 
m_A^*=\left [\frac{1}{m_A}+2V_A^1 \right ]^{-1},
\end{eqnarray} 
where $p^2V_{\tau }^1$ and $p^2V_A^1$ are the momentum dependent contributions
to the  single-particle potentials 
$V_\tau $ and $V_A$ :
\begin{eqnarray} 
V_\tau (p)=V_\tau^0+p^2V_\tau^1 ,
\end{eqnarray} 
\begin{eqnarray} 
V_A (p)=V_A^0+p^2V_A^1 .
\end{eqnarray} 
Expressions for the momentum independent components $V_\tau^0$ and 
$V_A^0$, alongwith those for $V_\tau^1, V_\tau^2$ and $V_A^1$ are
given in the Appendix.

\subsection{Energy and free energy of the system}

We take recourse to liquid-drop model for the evaluation of the total energy
$E_A(\rho ,T)$ of the fragments of mass $A$, charge $Z$ and neutron number $N$
at a constant density $\rho$ and temperature
$T$. The energy $E_A(\rho ,T)$ (=$-B_A(\rho ,T)$) is given by
\begin{eqnarray}
E_A(\rho,T)&=&a_v(\rho,T)A+a_s(\rho,T)4\pi R_A^2A^{2/3} \nonumber \\
&&+a_{sym}(\rho,T)\frac {(N-Z)^2}{A} \nonumber \\ 
&&+\frac {3}{5} {Z^2e^2} \Bigl (\frac{1}{R_A}-\frac{1}{R_{WS}} \Bigr )~.
\end{eqnarray} 
The term $a_v$ is the volume energy term for symmetric nuclear matter.
Alongwith $a_v$, the surface energy coefficient $a_s$ and the symmetry
energy coefficient $a_{sym}$ are all density and temperature dependent.
The last term in Eq.~(24) is the the Coulomb term. 
One may note that the Coulomb energy is different from that for
an isolated nucleus. As the fragment is embedded in clusterized 
matter, its Coulomb energy  gets 'dressed'. It is calculated
in the Wigner-Seitz approximation \cite{bon2}. Here $R_{WS}$
is the radius of the spherical Wigner-Seitz cell, given as
$R_{WS}= (\frac{4}{3}\pi \rho_A)^{-1/3}$. The Coulomb energy has no
explicit temperature dependence.
The radius $R_A$ of the liquid drop is given by
$R_A=A^{1/3}/[\frac{4}{3}\pi \rho(T)]^{1/3}$. In a similar vein to Eq.~(24), 
the free energy
of the nucleus is taken as 
\begin{eqnarray}
F_A(\rho,T)&=&f_v(\rho,T)A+f_s(\rho,T)4\pi R_A^2A^{2/3} \nonumber \\
&&+f_{sym}(\rho,T)\frac {(N-Z)^2}{A} \nonumber \\
&&+\frac {3}{5} {Z^2e^2} \Bigl (\frac{1}{R_A}-\frac{1}{R_{WS}} \Bigr )~.
\end{eqnarray} 
The volume terms $a_v$ and $f_v$ are calculated 
for symmetric nuclear matter at density
$\rho $ and at temperature $T$ employing the SBM interaction.
The density and temperature dependence of the surface free energy
coefficient is assumed to be factorized \cite{rav} and is taken as
\begin{eqnarray}
f_s(\rho ,T)=a_s(\rho_0,T=0){\cal U}(\rho ){\cal Y}(T),
\end{eqnarray} 
where $a_s(\rho_0 ,T=0)$ is the surface energy coefficient at nuclear matter
saturation density $\rho_0$ at $T=$0. The expressions for ${\cal U}(\rho )$
and ${\cal Y}(T)$ are taken from Refs. \cite{bla} and \cite{bon2},
respectively. They are given as 
\begin{eqnarray}
{\cal U}(\rho )=1-\frac {k_{\rho }}{2}\Bigl (\frac {\rho -\rho_0}{\rho_0}
\Bigr )^2,
\end{eqnarray} 
and
\begin{eqnarray}
{\cal Y}(T)= \Bigl (\frac {T_c^2-T^2}{T_c^2+T^2} \Bigr )^{5/4}.
\end{eqnarray} 
$T_c$ is the critical temperature for nuclear matter calculated to 
be 14.9~MeV with the SBM interaction.
 The value
of $a_s(\rho_o,T=0)$ and $k_{\rho}$ are taken to be 1.15~MeVfm$^{-2}$ and
5.0, respectively. The surface entropy per unit area ${\cal S}_{surf}$ is obtained
from $f_s$ as
\begin{eqnarray}
{\cal S}_{surf}=-\frac {\partial f_s}{\partial T}\Big |_{\rho}~~,
\end{eqnarray} 
which yields
\begin{eqnarray}
&&a_s(\rho ,T)=f_s(\rho ,T)+T{\cal S}_{surf} \nonumber \\
&&=a_s(\rho_0,0)\Bigl [{\cal Y}(T) +5 \Bigl (\frac{T_c^2-T^2}
{T_c^2+T^2} \Bigr )^{1/4} \nonumber \\
&&\times \frac {T_c^2T^2}{(T_c^2+T^2)^2}
\Bigr ]{\cal U}(\rho).
\end{eqnarray} 

The symmetry coefficient $a_{sym}$ is dependent on the nuclear mass. It
is taken as \cite{dan}
\begin{eqnarray}
a_{sym}(\rho =\rho_0,T=0)=\frac {\alpha}{1+\frac{\alpha}{\beta }A^{-1/3}},
\end{eqnarray} 
where $\alpha $ is the symmetry coefficient of cold symmetric nuclear
matter taken as 31.0 MeV and $\frac{\alpha }{\beta }=$2.4. For infinite matter,
it is generally seen that $a_{sym}$ decreases with temperature whereas
$f_{sym}$ shows the opposite temperature dependence \cite{xu}. A nearly similar
trend has been observed for finite nuclei \cite{de5};
here $f_{sym}$ increases with temperature (though in some cases,
an occasional decrease is seen at low $T$).
The density dependence of the symmetry energy of nuclear matter calculated
with the SBM interaction is seen to be given by  
$\sim (\rho/\rho_0)^\gamma $ with $\gamma \sim ~$0.69 \cite{sam1}, 
in consonance with the reported experimental behavior \cite{she1}.
With this in mind, we write 
the symmetry free energy coefficient in a factorized form as
\begin{eqnarray}
f_{sym}(\rho ,T)=a_{sym}(\rho, T=0 )g(T),
\end{eqnarray} 
where 
\begin{eqnarray}
a_{sym}(\rho,T=0 ) = a_{sym}(\rho_0,T=0 )(\rho /\rho_0 )^\gamma.
\end{eqnarray}
For $g(T)$, we assume a polynomial in $T$ of the form,
\begin{eqnarray}
g(T)=(1+\nu_1T+\nu_2T^2+\nu_4T^4).
\end{eqnarray}
Then,
\begin{equation}
{\cal S}_{sym}=-\frac {\partial f_{sym}}{\partial T}\Big |_{\rho}~~,
\end{equation}
 and therefore
\begin{eqnarray}
a_{sym}(\rho ,T)=f_{sym}(\rho ,T)-T\frac{\partial f_{sym}(\rho ,T)}
{\partial T} \Bigg |_{\rho }
\end{eqnarray}
\begin{eqnarray}
=a_{sym}(\rho,T=0 )[1-\nu_2T^2-3\nu_4T^4].
\end{eqnarray}
In a schematic model \cite{de5}, the observed $T$- dependence of $a_{sym}$
and $f_{sym}$ has been seen to be moderately 
explained with values of $\nu_1 ,\nu_2 ,$
and $\nu_4$ as $-$0.00848, 0.00201 and 0.0000147, respectively,
the dimensions of these quatities being in relevant inverse powers of MeV. 

The internal entropy $S_A^{int}$ 
of the fragments has contributions from the volume, surface, and
the asymmetry. The latter two contributions have already been taken
into account through Eqs.~(29) and (35), respectively. Since
the fragments are taken to have uniform density $\rho $, the volume entropy is
calculated using the expression given in  Eq.~(15) for symmetric
nuclear matter at  temperature $T$ and at a density $\rho $.
  
Combining the terms given by Eqs.~(7)-(10), the total energy of the
(n,p,A) system can then be written as,
\begin{eqnarray}
E&=&V \Biggl [ \Biggl \{ \sum_\tau \rho_\tau T J_{3/2}(\eta_\tau )
/J_{1/2}(\eta_\tau ) \nonumber \\
&& \times (1-m_{\tau }^*V_{\tau }^*) +\frac {1}{2}\rho_\tau V_{\tau }^0 
\Biggr \} 
 +\rho_A TC_A \nonumber \\
&& \times (1-m_A^*V_A^1)+ \frac{1}{2} \rho_AV_A^0-\rho_AB_A(
\rho_l ,T) \Biggr ] ~.
\end{eqnarray}
We may remind here that
$\rho_\tau $ corresponds to the free nucleonic density after condensation.
The total entropy of the free nucleonic matter is, from Eq.~(15),
\begin{eqnarray}
\sum_\tau S_\tau =V\sum_\tau \rho_\tau  
\Bigr [ \frac{5}{3}J_{3/2}(\eta_\tau )/
J_{1/2}(\eta_\tau )-\eta_\tau \Bigr ].
\end{eqnarray}
Similarly, the translational entropy from the fragments is,
\begin{eqnarray}
S_A^{tr}=N_A[\frac{5}{3}C_A-\eta_A],
\end{eqnarray}
where $C_A$ is given by,
\begin{eqnarray}
C_A= J_{3/2}(\eta_A)/J_{1/2}(\eta_A),
\end{eqnarray}
or
\begin{eqnarray}
C_A= B_{3/2}(\eta_A)/B_{1/2}(\eta_A),
\end{eqnarray}
depending on whether the fragments are fermionic or bosonic.
 In the above equations, the quantities $J_k$ and $B_k$ are the
Fermi and Bose integrals; their definitions are given in Eq.~(A.4)
in the Appendix A.
As the fragment densities are usually very low, $C_A \sim $3/2.
Since the internal entropy $S_A^{int}$ of the fragments
is now known as explained earlier, the free energy $\cal{F}$
of the total AN matter can be calculated.

\section{Results and Discussions}

In this paper, our primary aim is to investigate the changes in the 
properties of nuclei embedded in a hot medium of nucleons and other
fragments produced in nuclear multifragmentation. To simplify the
problem yet retaining the main physics essence, we assume that after nuclear
disassembly, the system contains a collection of 
only one kind of fragments of mass
$A$   and charge $Z$ in thermodynamic equilibrium, with a 
hot soup of neutrons and protons.  To begin with, 
we take a baryon matter of given density $\rho_b$, at a temperature
$T$ with an isospin asymmetry $X$. The binding energies and the 
free energies of the nuclear fragments that enter into the calculation
have been modeled in the  context of the liquid drop mass formula.
For the effective interaction, the momentum and density dependent
SBM force as scripted in Eqs.~(1) and (2) has been chosen. 
Assumptions are made that the free nucleons do not penetrate the 
sharp-surface nuclei and that the fragments do not overlap.
\begin{figure}
\includegraphics[width=1.0\columnwidth,angle=0,clip=true]{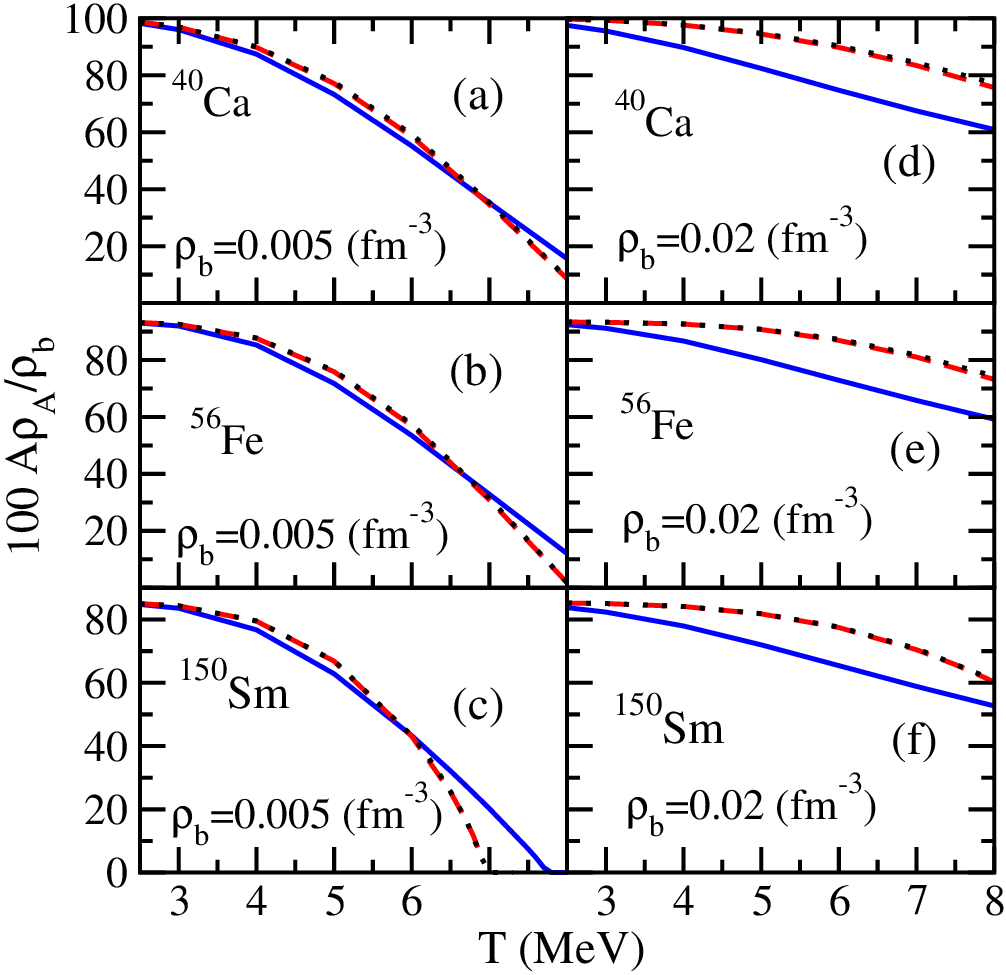}
\caption{ (Color online) The percentage of nucleons 
contained in the fragments at
a given baryon density $\rho_b$ 
at $X=$ 0.0 shown as a function of temperature
after fragmentation. The left panels correspond to $\rho_b=$ 0.005 fm$^{-3}$,
the right panels to $\rho_b=$ 0.02 fm$^{-3}$. The fragment specimens 
chosen are $^{40}$Ca, $^{56}$Fe and $^{150}$Sm, respectively.
The blue full line corresponds to the case (1,1,1),  i.e., the calculation
where all three interactions NN, AN, and AA are included.
The red dashed line refers to the case (1,0,1) where the AN
contribution is neglected, whereas the black dotted line corresponds
to the (1,0,0) calculation without both the AN and AA contributions.
For details, see text.}
\end{figure}
\begin{figure}
\includegraphics[width=1.0\columnwidth,angle=0,clip=true]{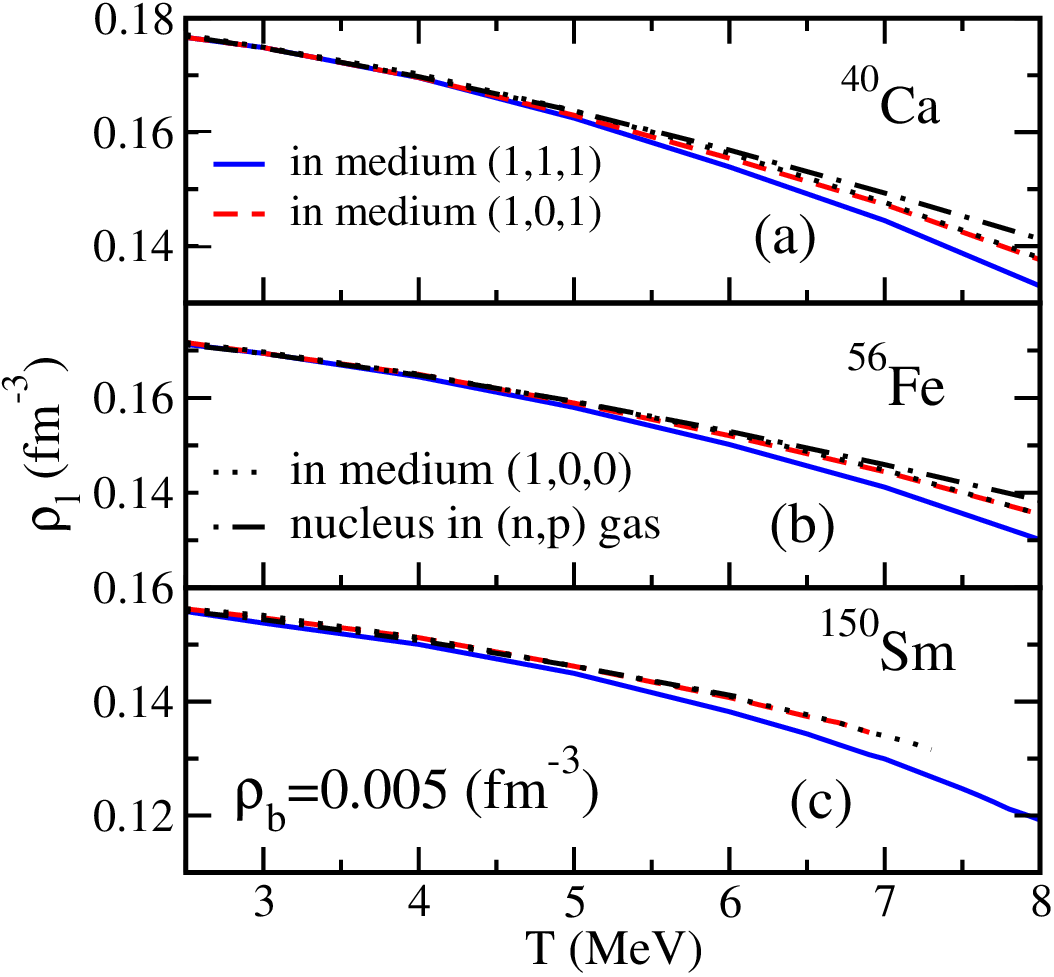}
\caption{ (Color online) The nucleon number density in the fragments
shown as a function of temperature, for baryon density $\rho_b=$ 0.005
fm$^{-3}$ at $X=$ 0.0. The blue full line, red dashed line and the
black dotted line convey the same meaning
as in Fig.~3. The dash-dot (black) lines refer to the nuclear
density when the nuclei ($^{40}$Ca, $^{56}$Fe or $^{150}$Sm)
are in phase equilibrium with their own vapor. }
\end{figure}

The three unknowns in the calculation are the free nucleon densities
$\rho_n, \rho_p $ and the fragment densities $\rho_A$ in the matter.
The three constraints are the conservation of the total baryon number,
the total isospin and the condition of chemical equilibrium between
the nucleon gas and the fragments. For a given set of $\rho_b, T $
and $X$, the calculations start with a chosen value of the
density $\rho_l$ of the constituent nucleons 
in the nuclear fragments at 
temperature $T$ and an input guess density of $\rho_A$. The energies
and the free energies of the fragments are then known from Eqs~(24)
and (25). Exploiting the constraints, the final densities $\rho_A,
\rho_n$ and $\rho_p$ are determined iteratively. The total free energy
$\cal {F}$ of the given $AN$ matter (see Eqs~(6) and (14)) is then calculated
as outlined earlier. Changing $\rho_l$ gives different values of $\cal {F}$,
whose minimum determines $\rho_l$ at a given 
baryonic density $\rho_b$ with asymmetry $X$
at temperature $T$, which then, in the liquid drop framework, determines
all the properties under investigation of the produced fragments.

The calculations have been done in the temperature range 2.5$-$8 MeV
for symmetric and asymmetric nuclear matter. 
Initially a symmetric matter ($X= 0.0$) of low baryon density 
$\rho_b =$ 0.005 fm$^{-3}$ is chosen,  
and the calculations are then repeated at a
higher baryon density $\rho_b$ =0.02 fm$^{-3}$. This helps 
to see how a denser medium
accentuates changes in the nuclear properties.  
Three representative fragments
are selected, namely, $^{40}$Ca, $^{56}$Fe and a heavier one $^{150}$Sm.
In Fig.~3, the percentage of nucleons in the fragments produced 
($A\rho_A/\rho_b \times $100) is shown as a function of temperature
for the three fragment species. The left panels correspond to $\rho_b$
=0.005 fm$^{-3}$, the right panels refer to the higher baryon density
$\rho_b$ =0.02 fm$^{-3}$. The blue full line is obtained from calculations
with inclusion of all three  interaction 
contributions, namely, (NN), (AN), and
(AA) [see Eqs.~(7)-(10)]. We refer to these calculations
as (1,1,1). The dashed red
line corresponds to calculations without the (AN) contribution and
the dotted black line is the one obtained when both (AN) and (AA)
contributions are excluded. The latter two calculations are referred to
(1,0,1) and (1,0,0), respectively. At low temperatures,
most of the nucleons are contained in the fragments and the free
nucleons are rare. This is expected, vapor tends to condense 
to drops at low temperature. 
As temperature increases, the fragment formation probability
decreases. Fragment formation also depends on 
the total  baryon density $\rho_b$;  
at the higher  $\rho_b$, fragment formation probability is higher.
As the system heats up, this probability goes down.

Examination of the figure reveals few further features. The nucleus-nucleus
interaction (AA) does not have a very significant role (as seen from
the almost overlapping of the red dashed (1,0,1) 
and black dotted (1,0,0) lines),
the nucleon-nucleus interaction (AN) is important, the importance
grows with increasing baryon density $\rho_b$. Normally, it is seen
that the (1,0,1) or (1,0,0) calculations favor the production of
fragments compared to a full (1,1,1) calculation.
These results allow us to conclude that heavy nuclei embedded in
the medium are affected mostly by the nucleons (and perhaps light
clusters) surrounding them. This means that the description of
the multicomponent nuclear system can be simplified by subdividing
it into non-interacting cells containing one heavy nucleus and 
a proportional amount of the medium. This Wigner-Seitz approximation
is widely used for studying inhomogeneous phases of nuclear
matter \cite{bonc,latt,maru}.

In Fig.~4, the progressive changes in the internal nucleonic density
in the three nuclei produced from disassembly of symmetric nuclear
matter of density $\rho_b$=0.005 fm$^{-3}$ are displayed as a function
of temperature in the three panels. The blue (full line), red (dashed)
and the black (dotted) lines have the same meaning as stated earlier.
The dash-dot black line refers to calculations for isolated hot
nuclei ($^{40}$Ca, $^{56}$Fe or $^{150}$Sm) in phase equilibrium
with their own vapor (n-p gas).
In that case, the phase-equilibrium conditions \cite{ban}
determine the density and asymmetry of the embedding nucleonic gas
alongwith the internal density of the dipped nucleus. The dot-dash
black line is drawn so as to serve as 
a reference against which the other ones can be 
compared.  In this case thermodynamic equilibrium ensures that the 
temperatures of 
the nucleus and the surrounding (n,p) gas are the same, the pressure
exerted by the nucleus balances that of the gas and that the chemical
potentials of the neutrons and protons of the nucleus are the same
as those of the free neutrons and protons in the gas, respectively.
Discussion on this part of physics is left out here, it is
given in detail in Ref. \cite{ban,lev}. The densities of the 
embedding n-p gas in clusterized matter and in the case of isolated
hot nucleus are somewhat different. 
In both cases
the density of the surrounding n-p gas is low at low temperatures
and increases with increase in temperature. The asymmetry of this
gas is also different in both cases. 
As an illustrative example, these properties of the embedding 
nucleonic gas are displayed in Fig.~5. The asymmetry of the disassembling
system is $X$ =0.2 and the fragment concerned is $^{150}$Sm. 
The asymmetry of the gas has a very insignificant role to play
on the properties of the fragments as will be shown later. 
\begin{figure}
\includegraphics[width=1.0\columnwidth,angle=0,clip=true]{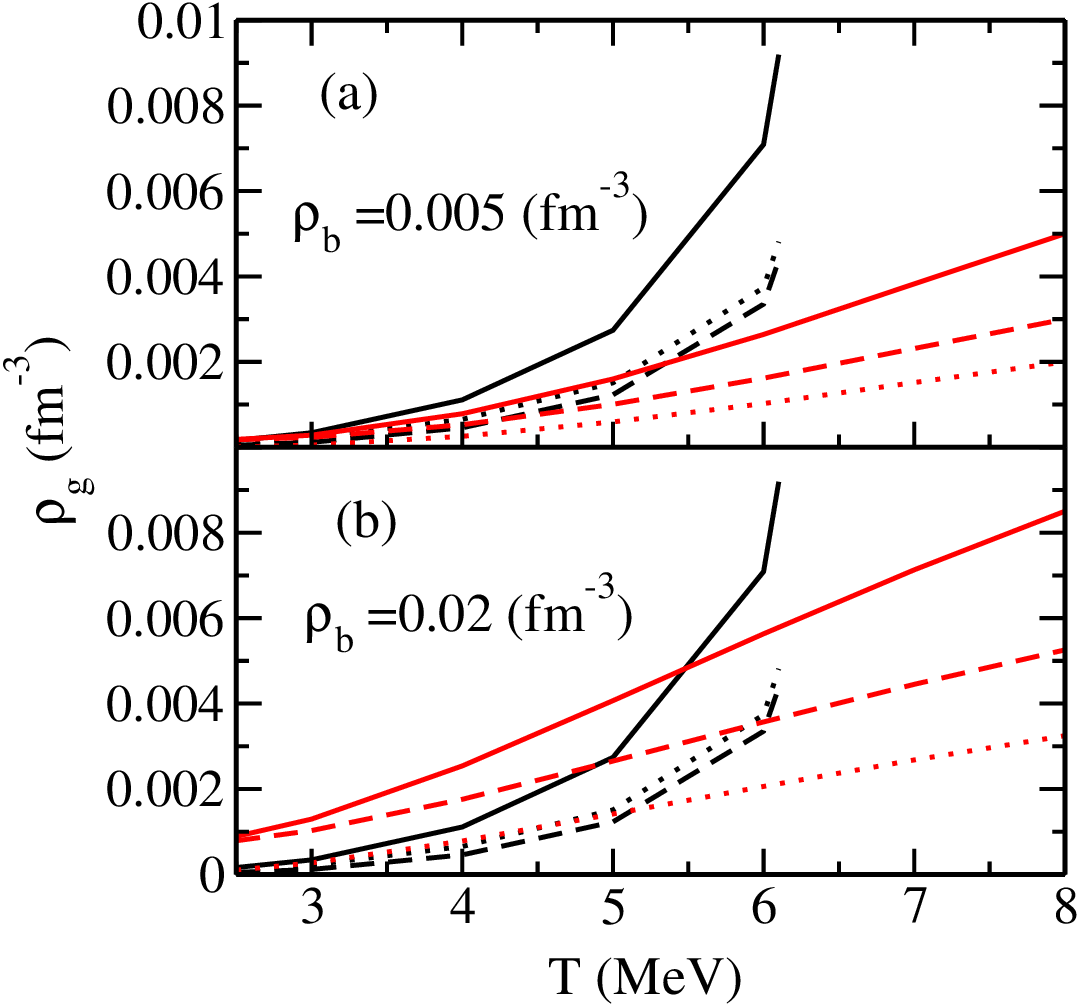}
\caption{ (Color online) The embedding nucleon gas density plotted
as a function of temperature in AN matter where A is $^{150}$Sm.
The panels (a) and (b) correspond to $\rho_b$ = 0.005 fm$^{-3}$
and 0.02 fm$^{-3}$, respectively. The red dotted, dashed and full lines
refer to proton and neutron densities and their sum in the
calculation for disassembled matter at asymmetry $X$ =0.2.
The black dotted, dashed and full lines correspond
to proton and neutron densities and their sum in the phase-equilibrium
calculation for $^{150}$Sm.}
\end{figure}

From  Fig.~4, it transpires as expected, the fragment nuclei in
equilibrium in an embedding medium expand with temperature in all
four calculations  displayed in this figure. In the reference calculation 
(black dash-dot line),
nuclei like $^{40}$Ca or $^{56}$Fe bloat up in volume by $\sim $
17-19$\% $ from their ground state equilibrium values when the
temperature is raised to $\sim $ 7$-$8 MeV. In the full (1,1,1) fragmentation
calculation for nuclear matter at $\rho_b$=0.005 fm$^{-3}$ they do 
so by $\sim $ 25 $\% $. As is seen from the figure, incorporation
of the nucleon-fragment (AN) interaction plays a significant role in
the expansion of the fragments; also it is seen, as stated
earlier, that the fragment-fragment interaction (AA) has little
effect. In panel (c) of Fig.~4, it is noticed that
the black dash-dot line does not extend 
\begin{figure}
\includegraphics[width=1.0\columnwidth,angle=0,clip=true]{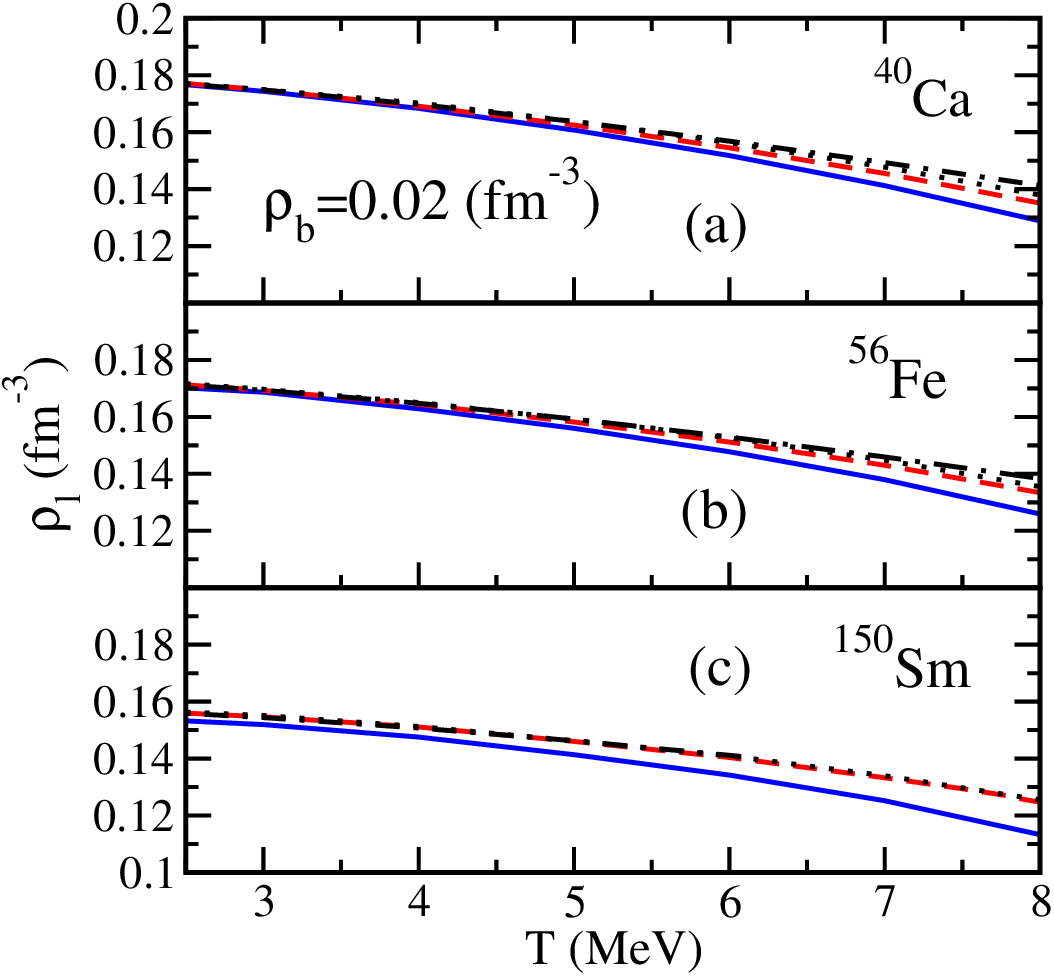}
\caption{ (Color online) Same as in Fig.~4 when $\rho_b=$ 0.02 fm$^{-3}$.}
\end{figure}
beyond 6.1 MeV. This is the limiting
temperature as obtained for  $^{150}$Sm 
in the phase equilibrium calculation.
\begin{figure}
\includegraphics[width=1.0\columnwidth,angle=0,clip=true]{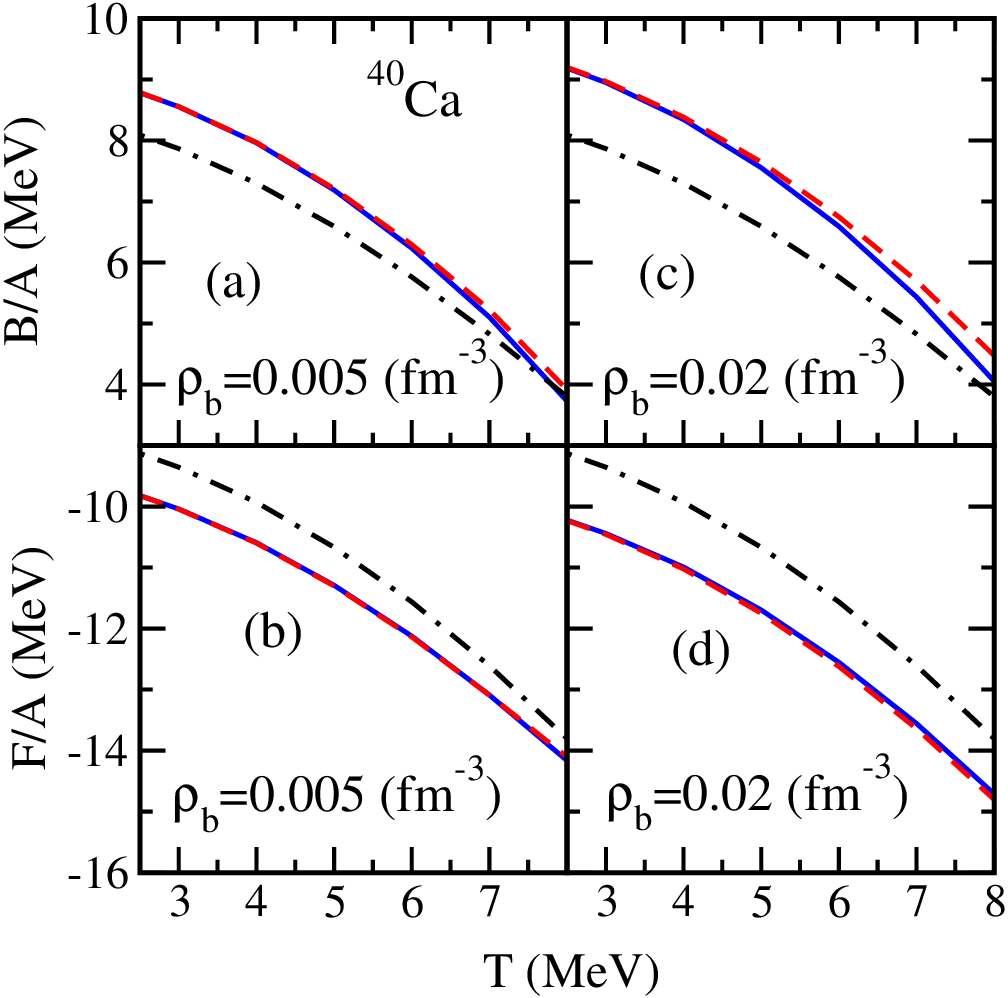}
\caption{ (Color online) The binding energy and free energy of
$^{40}$Ca, produced in fragmentation of symmetric nuclear matter. The
left panels refer to $\rho_b=$ 0.005 fm$^{-3}$, the right panels
correspond to $\rho_b=$ 0.02 fm$^{-3}$. The full (blue), dashed (red)
and the dash-dot (black) lines have the same meaning as in Fig.~4. }
\end{figure}
\begin{figure}
\includegraphics[width=1.0\columnwidth,angle=0,clip=true]{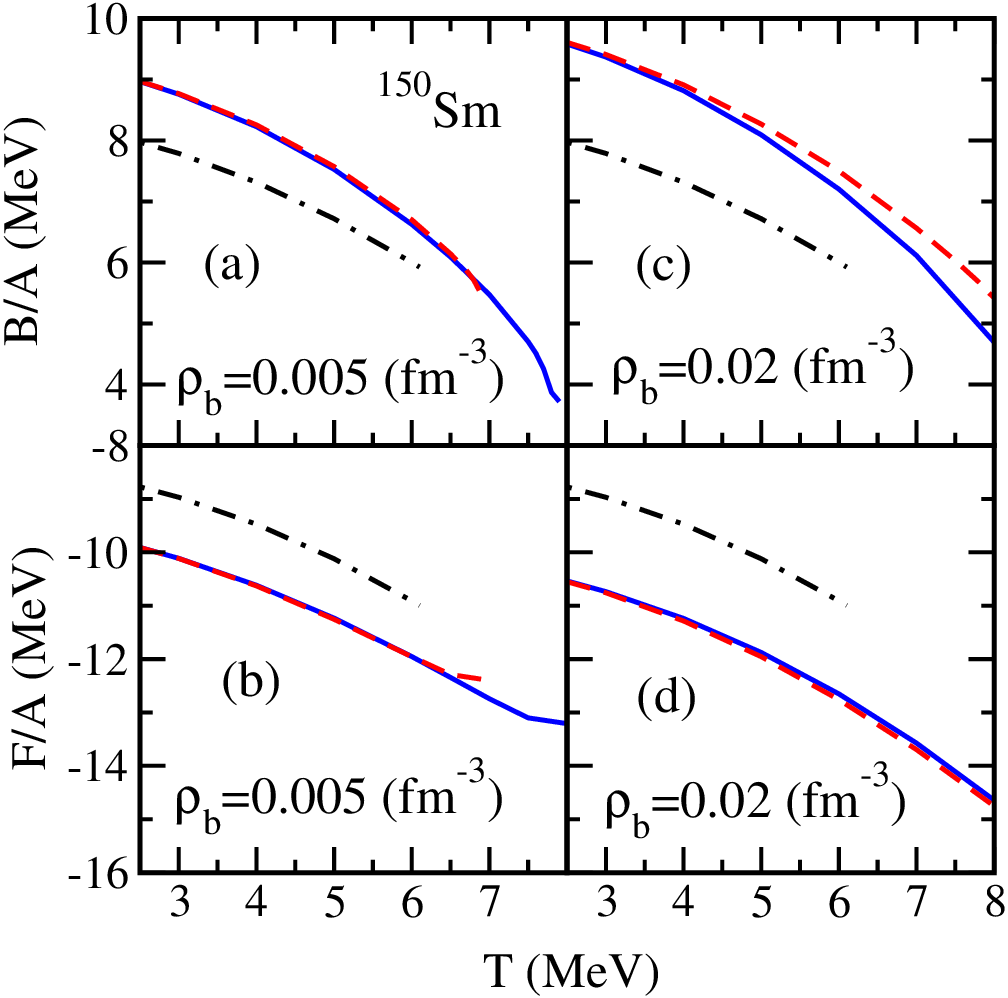}
\caption{ (Color online) Same as in Fig.~7 for $^{150}$Sm.}
\end{figure}
Solutions for calculations without incorporation
of both AN and AA (1,0,0) interactions and for calculations without
AN interactions (1,0,1) could not be obtained for this nucleus
beyond 7.2 and 6.9 MeV, repectively.
\begin{figure}
\includegraphics[width=1.0\columnwidth,angle=0,clip=true]{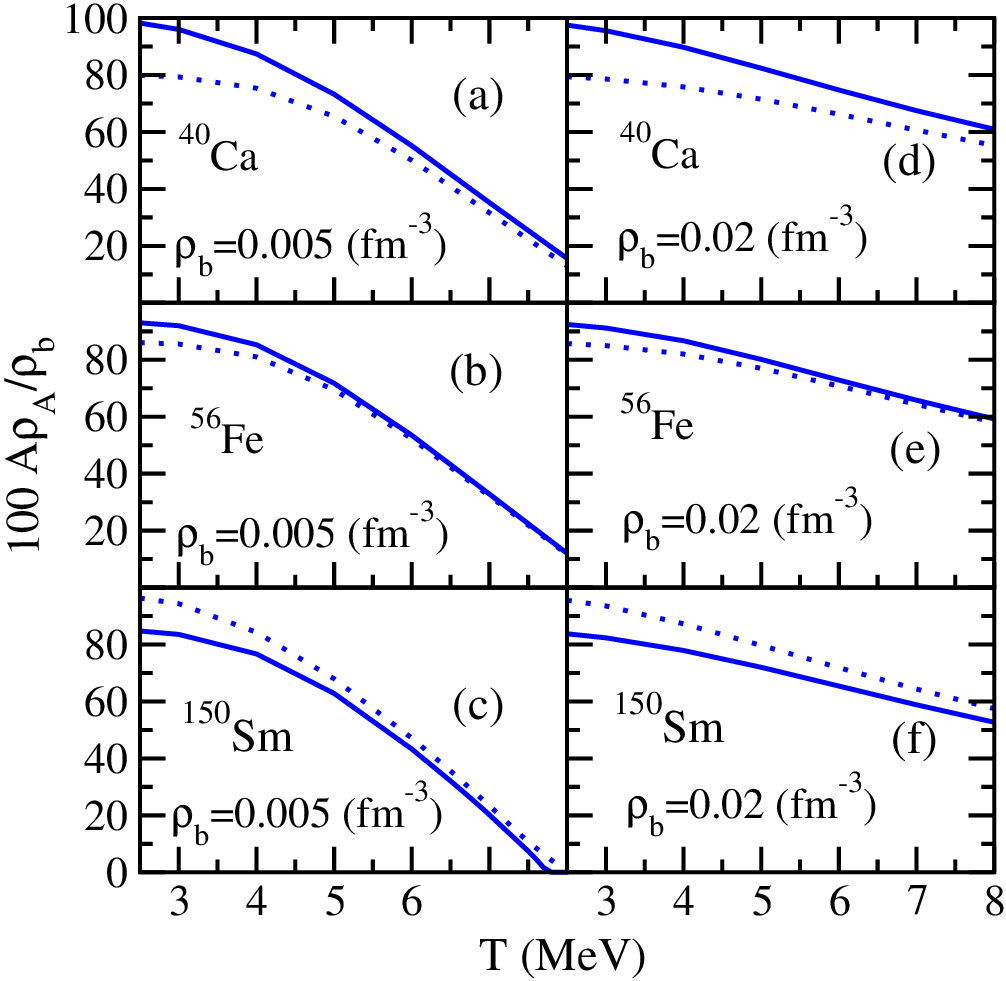}
\caption{ (Color online) Comparison of fragment population in 
symmetric and asymmetric  nuclear matter for the case (1,1,1)
calculated for two baryon densities. 
The full and dotted lines correspond
to $X =0.0$ and $X =0.2$, respectively.  }
\end{figure}
The so-found absence of solutions possibly
points to the nuclear instability beyond these temperatures
in these calculations. 
When the baryon density $\rho_b$ is increased, the nuclear
soup of fragments and nucleons becomes denser. 
\begin{figure}
\includegraphics[width=1.0\columnwidth,angle=0,clip=true]{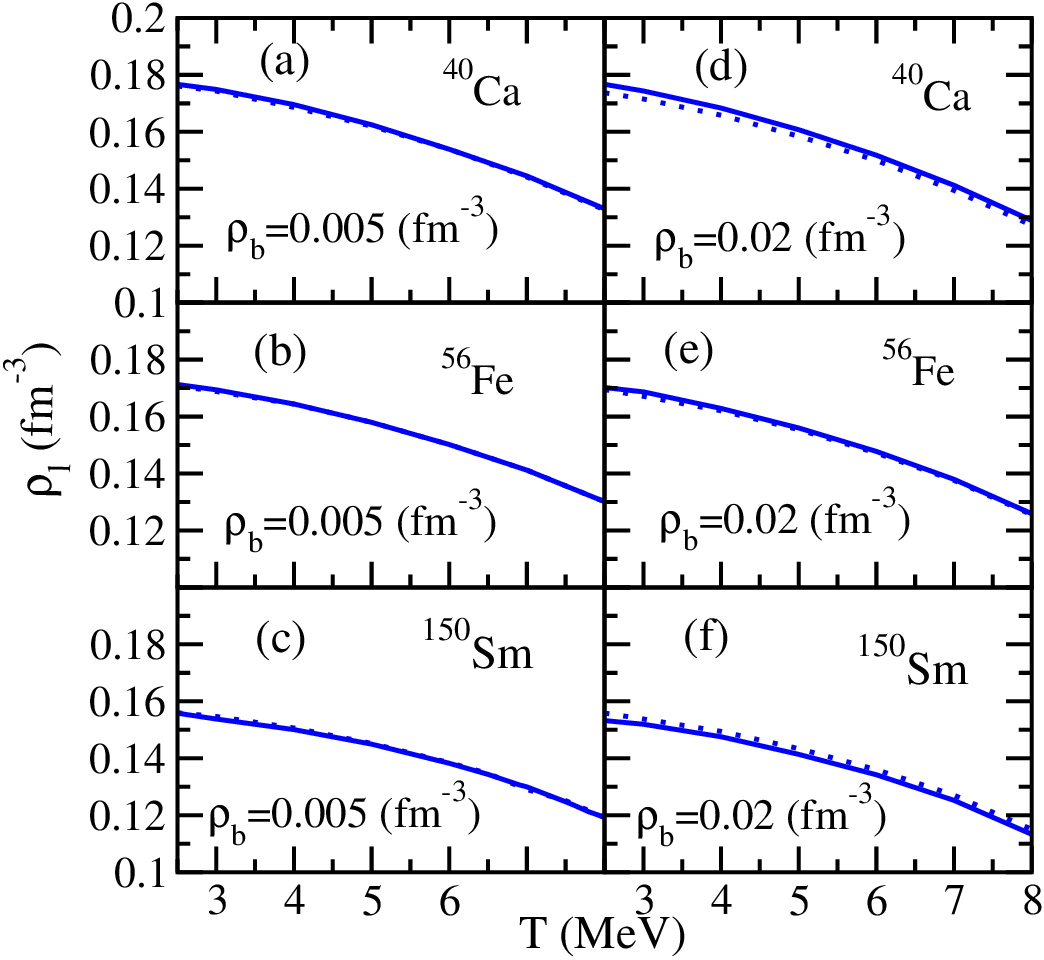}
\caption{ (Color online) Comparison of the internal nucleon densities
of the fragments produced in symmetric and asymmetric ($X =0.2$) 
nuclear matter calculated for two baryon densities. 
The full and dotted lines have the same meaning
as in Fig.~9.} 
\end{figure}
\begin{figure}
\includegraphics[width=1.0\columnwidth,angle=0,clip=true]{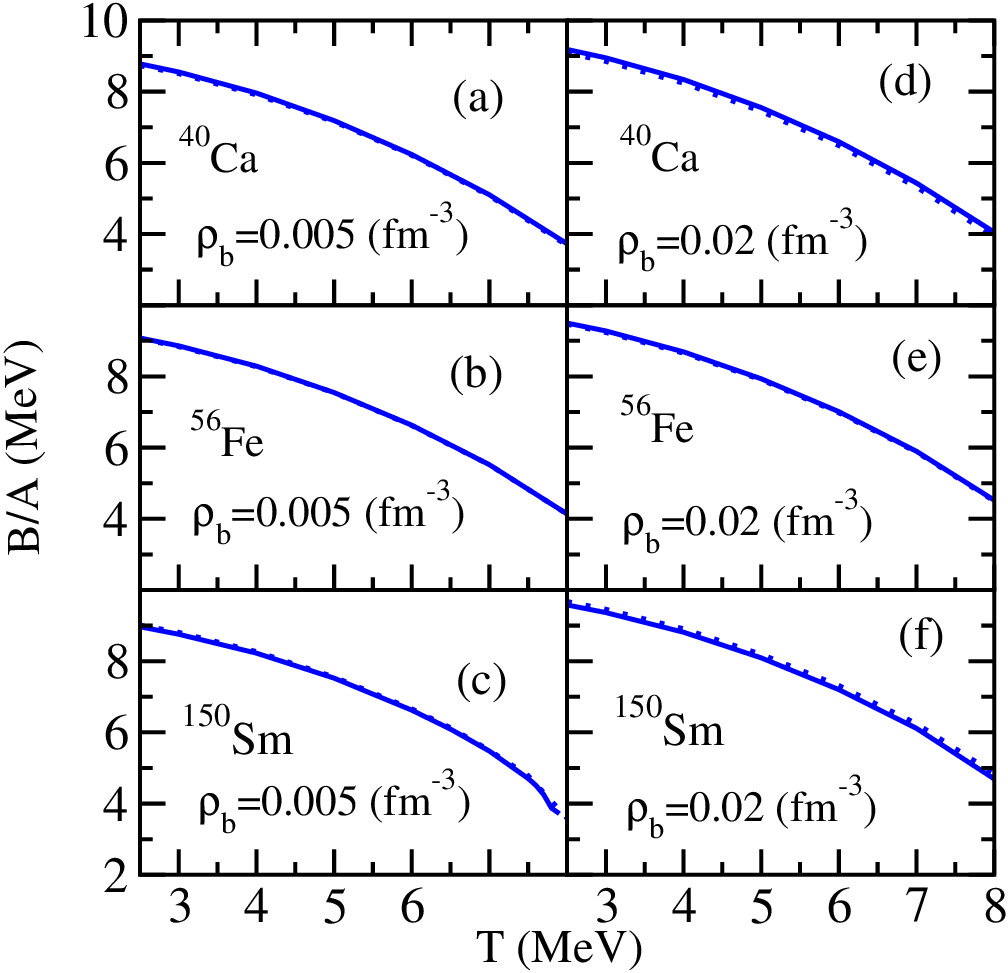}
\caption{ (Color online) Comparison of the binding energies 
of the fragments produced in symmetric and asymmetric ($X =0.2$) 
nuclear matter calculated for two baryon densities. 
The full and dotted lines have the same meaning
as in Fig.~9.} 
\end{figure}
In this denser environment, a further expansion
of the fragments, by another $\sim $4$\% $ could be noticed. This
is displayed in Fig.~6 where $\rho_b$ = 0.02 fm$^{-3}$. The (1,0,1)
and (1,0,0) calculations for $^{150}$Sm could now be extended to 8.0 MeV. 

As the internal density of the nuclei change progressively with temperature
from those obtained from the reference calculation as mentioned earlier,
it is expected that there should be a corresponding change in the binding
energies and free energies of the nuclei embedded in medium. We display
those quantities for $^{40}$Ca in Fig.~7.
 The left panels (a) and (b) show the results  
for the lower baryon density $\rho_b$=0.005 fm$^{-3}$, the
right panels (c) and (d) do so for $\rho_b$=0.02 fm$^{-3}$,
both at $X$ =0.0. The
upper panels display the binding energy, the lower panels the free energy.
The (1,0,0) calculations being nearly indistinguishable from the
(1,0,1) calculations are not shown here. 
The results for the binding energies from the (1,1,1) and (1,0,1)
calculations at the lower baryon density are nearly indistinguishable
from each other, at the higher baryon density 
 ($\rho_b$=0.02 fm$^{-3}$)
only little changes
are observed at the high temperatures. Compared to the reference
calculation (dash-dot black line representing phase equilibrium
in a n-p gas), there is a gain in the binding energy. This comes
mostly from the decrease of the Coulomb energy because of the presence
of other fragments (see Eq. 24). At higher baryon density, the fragment
density is comparatively higher which explains the larger gap
in binding energy with reference to the phase equilibrium 
calculation.  For a given baryon density the reduction in 
the density of the fragments 
with temperature reduces the gap. Similar arguments follow 
for the lower free energy of the embedded fragments as compared
to that from the reference calculation. Not much of a difference
is seen for the heavier nucleus $^{150}$Sm as displayed in Fig.~8.  

In order to see the importance of asymmetry on the fragment
observables, the calculations have been repeated for asymmetric
nuclear matter with $X=0.2$. In Fig.~9, comparison of the percentage
of nucleons contained in the fragments are made for symmetric
and asymmetric nuclear matter at the two baryon densities we have
considered for the three fragment species. We display only the
full calculations (1,1,1). The blue full lines correspond to
$X = 0.0$, the blue dotted lines refer to $X =0.2$. $^{40}$Ca
and $^{56}$Fe being symmetric and nearly symmetric nuclei,
respectively have comparatively larger population in symmetric
nuclear matter. For the more asymmetric $^{150}$Sm nucleus,
population is larger in the asymmetric matter. The difference
between the two calculations at the two asymmetries is
more pronounced at lower temperatures gradually narrowing down 
as the temperature is raised. The internal nucleon density of the
fragments and their binding energies, however, show no significant
change when the isospin asymmetry of the matter changes. This
is displayed in Fig.~10 and Fig.~11, respectively for the three
nuclear fragments at the two baryon densities; the blue full and 
dotted lines nearly overlap each other over the whole temperature
range we work in.

\section{concluding remarks}

Analysis of nuclear multifragmentation data showed that for the
results calculated in thermodynamic models 
to conform to the experimentally observed ones, the established 
nuclear parameters taken as inputs in these calculations 
needed subtle changes \cite{sou,ogu,bot2,buy}.
 Such a fact points
out that the properties of the fragments produced in nuclear disassembly
might have got modified because of the interaction of the fragments 
with the embedding environment they are created in. 
The calculations presented in this paper throw light in
a quantitative manner on how significant these modifications can be. 
For simplicity, the fragmented system
was assumed to contain a collection of only one kind of nuclear species
in addition to neutrons and protons. For comparison, a benchmark
calculation of the hot fragments in phase-equilibrium with its own 
vapor was also done. In both cases, it was seen that the fragments 
expand with temperature as expected, but compared to the results
from the phase-equilibrium (benchmark) calculation, the embedding
medium in multifragmentation produced larger changes in the fragment
properties. The fragments get comparably more stretched; the volume
energy, surface properties or the symmetry properties of the fragments
undergo the consequential changes. 

Questions may arise on the justification of the choice of only one kind
of species in the calculations. Close examination of the results shows
that in the medium, the interaction of the nucleons with the fragments plays
the dominant role in bringing forth the modification in the fragment
properties. The fragment-fragment interaction has a very nominal role.
The selection of the multispecies in the medium may not thus possibly
alter the results much.

Experiments with intermediate energy heavy ion beams in the last
few decades have indicated \cite{aug} that nuclei can sustain only
temperatures that are much lower than the critical temperature
($\sim $ 16 MeV) for symmetric nuclear matter. The origin of such
a limiting temperature is usually traced to an interplay between 
the Coulomb instability and the  corrections due to
the finite size of the nuclear drops. Phase-equilibrium calculations 
for hot isolated nuclei surrounded by their own vapor yield limiting
temperatures $\sim $5$-$7 MeV for heavier nuclei \cite{lev,sur,ban}.
In the calculations presented in this paper, it is seen that for
nuclei dripped in a nuclear soup, the interaction with the surrounding
medium might overcome the said instability to a certain extent
and extend somewhat the limit in temperature the nuclei may hold. 
This might be of 
significant relevance in the context of nuclear astrophysics 
and needs further exploration.

\begin{acknowledgments}
J.N.D and S.K.S acknowledge the support of DST, Government of India.
M.C. and X.V. acknowledge the support of the Consolider Ingenio 2010
Programme CPAN CSD2007-00042, of the grants 
FIS2011-24154 from MICINN and FEDER, and of grant 2009SGR-1289 from
Generalitat de Catalunya. I.N.M acknowledges support from grant
NSH-215.2012.2, Russia.
\end{acknowledgments}

\appendix
\section{}
\setcounter{equation}{0}
\renewcommand{\theequation}{A.\arabic{equation}}

 From Eqs.~(11) and (18), the density of the free nucleons $\rho_\tau$
is found to be
\begin{eqnarray}
\rho_\tau=\frac{4\pi}{h^3}(2m_\tau^*T)^{3/2}J_{1/2}(\eta_\tau).
\end{eqnarray}
Similarly, from Eqs.~(12) and (19), the fragment density $\rho_A$
comes out as
\begin{eqnarray}
\rho_A=\frac{4\pi}{h^3}(2m_A^*T)^{3/2}J_{1/2}(\eta_A),
\end{eqnarray}
or
\begin{eqnarray}
\rho_A=\frac{2\pi}{h^3}(2m_A^*T)^{3/2}B_{1/2}(\eta_A),
\end{eqnarray}
depending on whether the fragments are fermions or bosons.
The $J_k(\eta)$ and  $B_k(\eta)$ are the Fermi and Bose integrals,
\begin{eqnarray}
J_k(\eta)=\int_0^\infty \frac{x^k dx}{e^{(x-\eta)}+1}, \nonumber \\
B_k(\eta)=\int_0^\infty \frac{x^k dx}{e^{(x-\eta)}-1}.
\end{eqnarray}

 The expression for $V_\tau^0$ in Eq.~(22) is given as,
\begin{eqnarray} 
&& V_\tau^0= -4\pi a^3\left \{ 1-d^2(2\rho_N )^\kappa \right \}
(C_l\rho_\tau +C_u\rho_{-\tau }) \nonumber \\
&& +\frac{16\pi^2a^3}{b^2h^3} \biggl [C_l(2m_\tau^*T)^{5/2}J_{3/2}
(\eta_\tau )+C_u(2m_{-\tau }^*T)^{5/2} \nonumber \\
&& \times J_{3/2}(\eta_{-\tau }) \biggr ] 
+\frac{1}{2}I\rho_A \biggl \{(\frac{<p_A^2>}{b^2}-1)(C_l\rho_{l\tau}
+C_u\rho_{l-\tau}) \nonumber \\
&& +\frac{4\pi}{h^3b^2}[C_l(2m_{l\tau}^{A*}T)^{5/2}J_{3/2}(\eta_{l\tau}^A)
+C_u(2m_{l-\tau}^{A*})^{5/2} \nonumber \\
&&\times J_{3/2}(\eta_{l-\tau}^A)] 
+(C_l\rho_{l\tau}+C_u\rho_{l-\tau})d^2[\rho_N+
\rho_l]^\kappa \biggr \} .
\end{eqnarray} 
In this equation, the  fugacity $\eta_{l\tau}^A$ is defined corresponding
to the nucleons of density $\rho_{l\tau }$ 
inside the fragment (exactly in parallel definition of the 
fugacity $\eta_\tau$ of the free nucleons corresponding
to the free nucleon density $\rho_\tau$); $m_{l\tau}^{A*}$
is the effective mass of these nucleons.

Similarly, expressions for $ V_{\tau }^1, V_{\tau }^2,
V_A^0$ and $V_A^1$ are, 
\begin{eqnarray} 
 V_\tau^1=\frac{4\pi a^3}{b^2}[C_l\rho_\tau +C_u \rho_{-\tau }]
+\frac{1}{4}I(C_l+C_u) \frac{\rho_A \rho_l }{b^2},
\end{eqnarray} 
\begin{eqnarray} 
V_\tau^2&=& 4 \pi a^3 \kappa d^2(2\rho_N )^{\kappa -1} \sum_{\tau \prime }
[C_l\rho_{\tau ^\prime }+C_u \rho_{-\tau ^\prime }]\rho_{\tau ^\prime }
\nonumber \\
&& +\frac{1}{2}I \rho_A \rho_N 
(C_l\rho_{l\tau}+C_u\rho_{l-\tau}) \nonumber \\
&&\times [\kappa d^2
(\rho_N+\rho_l)^{\kappa -1}]~,
\end{eqnarray} 
\begin{eqnarray} 
V_A^0&=&\frac{1}{2}I\sum_\tau \rho_\tau \Bigl [ (C_l\rho_{l\tau}
+C_u\rho_{l-\tau}) 
 \Bigl \{-1+\frac{4\pi}{h^3}\frac{(2m_\tau ^*T)^{5/2}}{b^2\rho_\tau }
\nonumber \\
&& \times J_{3/2}(\eta_\tau )+d^2\{\rho_N
+\rho_l\}^\kappa
\Bigr \} \nonumber \\
&&+\frac{4\pi}{b^2h^3}\Bigl \{C_l(2m_{l\tau}^{A*}T)^{5/2}
J_{3/2}(\eta_{l\tau }^A) \nonumber \\
&&+C_u(2m_{l-\tau}^{A*}T)^{5/2}J_{3/2}(\eta_{l-\tau}^A) \Bigr \}\Bigr ] 
\nonumber \\
&&+I_A\rho_A \sum_\tau \rho_{l\tau} \Biggl \{C_l \rho_{l\tau}
\Bigl (-1+\frac{<p_A^2>}{b^2}+\frac{8\pi}{b^2h^3} \nonumber \\
&& \times \frac{(2m_{l\tau}^{A*}T)^{5/2}}
{\rho_{l\tau}}J_{3/2}(\eta_{l\tau}^A) \Bigr )
+C_u\rho_{l-\tau} \Bigl (-1+\frac{<p_A^2>}{b^2} \nonumber \\
&&+\frac{4\pi}{b^2h^3}\frac{(2m_{l\tau}^{A*}T)^{5/2}}{\rho_{l\tau}}
J_{3/2}(\eta_{l\tau}^A) \nonumber \\
&&+\frac{4\pi}{b^2h^3}\frac{(2m_{l-\tau}^{A*}T)^{5/2}}{\rho_{l-\tau}}
J_{3/2}(\eta_{l-\tau}^A) \Bigr ) \Biggr \} \nonumber \\
&& +I_A \rho_Ad^2\Bigl [\sum_\tau \rho_{l\tau} (C_l\rho_{l\tau}
+C_u\rho_{l-\tau})\Bigr ](2\rho_l)^\kappa ~,
\end{eqnarray} 
\begin{eqnarray} 
V_A^1=\sum_\tau  (C_l\rho_{l\tau}+C_u\rho_{l-\tau})
(\frac{I_A\rho_A}{b^2}\rho_{l\tau}+\frac{I}{2b^2}\rho_\tau).
\end{eqnarray} 
In Eqs.~(A.5)-(A.8), $<p_A^2>$ is the mean squared value of the
fragment momentum in AN matter. Its value is given by,
\begin{eqnarray}
<p_A^2>=(2m_A^{*}T)C_A,
\end{eqnarray}
where $C_A$ is given by Eqs.~(40) or (41).
The integrals $I$ and $I_A $ appearing in Eq.~(A.5)-(A.9)
for nucleon-nucleus and nucleus-nucleus interactions are given by,
\begin{eqnarray} 
I=\int_{V_A} d{\bf r} \int d{\bf R} \frac{e^{-|{\bf r}+{\bf R}|/a}}
{|{\bf r}+{\bf R}|/a},
\end{eqnarray} 
\begin{eqnarray} 
I_A=\int_{V_A }d{\bf r }\int_{V_A }d{\bf r^\prime }
\int d{\bf R} \frac{e^{-|{\bf R}+{\bf r}-{\bf r^\prime }|/a}}
{|{\bf R}+{\bf r}-{\bf r^\prime }|/a},
\end{eqnarray} 
Integrations on ${\bf R }$ exclude the fragment volumes. 
The integrals are evaluated numerically.

\end{document}